\documentclass[prd,preprint,nofootinbib]{revtex4}

\usepackage{graphicx}
\usepackage{amssymb}
\usepackage{amsmath}
\usepackage{color}

\begin{document}
\renewcommand{\thefootnote}{\#\arabic{footnote}}
\newcommand{\rem}[1]{{\bf [#1]}}
\newcommand{\gsim}{ \mathop{}_ {\textstyle \sim}^{\textstyle >} }
\newcommand{\lsim}{ \mathop{}_ {\textstyle \sim}^{\textstyle <} }
\newcommand{\vev}[1]{ \left\langle {#1}  \right\rangle }
\newcommand{\bear}{\begin{array}}  
\newcommand {\eear}{\end{array}}
\newcommand{\bea}{\begin{eqnarray}}   
\newcommand{\eea}{\end{eqnarray}}
\newcommand{\beq}{\begin{equation}}   
\newcommand{\eeq}{\end{equation}}
\newcommand{\bef}{\begin{figure}}  
\newcommand {\eef}{\end{figure}}
\newcommand{\bec}{\begin{center}} 
\newcommand {\eec}{\end{center}}
\newcommand{\non}{\nonumber}  
\newcommand {\eqn}[1]{\beq {#1}\eeq}
\newcommand{\la}{\left\langle}  
\newcommand{\ra}{\right\rangle}
\newcommand{\ds}{\displaystyle}
\newcommand{\red}{\textcolor{red}}
\def\SEC#1{Sec.~\ref{#1}}
\def\FIG#1{Fig.~\ref{#1}}
\def\EQ#1{Eq.~(\ref{#1})}
\def\EQS#1{Eqs.~(\ref{#1})}
\def\lrf#1#2{ \left(\frac{#1}{#2}\right)}
\def\lrfp#1#2#3{ \left(\frac{#1}{#2} \right)^{#3}}
\def\GEV#1{10^{#1}{\rm\,GeV}}
\def\MEV#1{10^{#1}{\rm\,MeV}}
\def\KEV#1{10^{#1}{\rm\,keV}}
\def\REF#1{(\ref{#1})}
\def\lrf#1#2{ \left(\frac{#1}{#2}\right)}
\def\lrfp#1#2#3{ \left(\frac{#1}{#2} \right)^{#3}}
\def\OG#1{ {\cal O}(#1){\rm\,GeV}}

\begin{flushright}
IPMU 08-0113\\
UT-08-34\\
KEK-TH 1294
\end{flushright}

\title{Dark Matter Model Selection and the ATIC/PPB-BETS anomaly}

\author{Chuan-Ren Chen$^{1}$, Koichi Hamaguchi${}^{1,2}$, Mihoko M. Nojiri${}^{1,3}$, \\
             Fuminobu Takahashi$^{1}$ and Shoji Torii${}^{4}$}

\affiliation{
${}^{1}$Institute for the Physics and Mathematics of the Universe,
University of Tokyo, Chiba 277-8568, Japan,\\
${}^{2}$Department of Physics, University of Tokyo, Tokyo 113-0033, Japan,\\
${}^{3}$ Theory Group, KEK and the Graduate University for Advanced Study,
	 Ibaraki, 305-0801, Japan,\\
${}^{4}$Research Institute for Science and Engineering, Waseda University, 3-4-1, Okubo,
Shinjuku-ku, Tokyo, 169-8555, Japan
}

\date{\today}

\begin{abstract}
We argue that we may be able to sort out dark matter models in which
electrons are generated through the annihilation and/or decay of dark
matter, by using a fact that the initial energy spectrum is reflected
in the cosmic-ray electron flux observed at the Earth even after
propagation through the galactic magnetic field.  To illustrate our
idea we focus on three representative initial spectra: (i)monochromatic
(ii)flat and (iii)double-peak ones.  We find that those three cases
result in significantly different energy spectra, which may be probed
by the Fermi satellite in operation or an up-coming cosmic-ray detector
such as CALET.
\end{abstract}


\maketitle

\section{Introduction}
\label{sec:1}
The presence of dark matter has been firmly established by numerous
observational data, although we have not yet understood what dark
matter is made of. Recent cosmic-ray measurements aiming for indirect dark matter
detection may be providing us with insights into dark matter.

The PAMELA data~\cite{Adriani:2008zr} showed that the positron
fraction starts to deviate from a theoretically expected value for
secondary positrons around $10$ GeV, and continues to increase up to
about $100$\,GeV. The ATIC
collaboration~\cite{ATIC-new} has recently released the data, showing
a clear excess in the total flux of electrons plus positrons peaked
around $600 - 700$\,GeV, in consistent with the PPB-BETS
observation~\cite{Torii:2008xu}.  The excess may be explained by
astrophysical sources like
pulsars~\cite{Aharonian(1995),Hooper:2008kg} or
microquasars~\cite{Heinz:2002qj}, although it is not easy to account
for the electron flux with a sharp drop-off observed by
ATIC~\footnote{ The nearby pulsars may be able to explain the PAMELA
  data, though~\cite{Hooper:2008kg}.  }.  An alternative explanation
is the annihilation and/or decay of dark matter.  Indeed, the exciting
PAMELA and ATIC/PPB-BETS data has stimulated new directions in dark-matter
model building~\cite{DM-models}. In this letter we 
take a step further toward sorting out those dark matter models.

One important constraint on the dark matter models comes from the
absence of any excess in the anti-proton
flux~\cite{Adriani:2008zq,Yamamoto:2008zz}.  Also, the PAMELA and
ATIC/PPB-BETS anomalies in the electrons and positrons suggest that
the initial energy spectrum of the electrons and positrons should be
hard.  Those observational evidences suggest that electrons (or muons)
must be directly produced from dark matter, with the hadronic branch
being suppressed.  The models proposed so far are broadly divided into
two categories concerning how to suppress the antiproton production.
One category is such that the dark matter particle mainly annihilates or decays
into leptons. For instance, the dark matter may be a hidden $U(1)_H$
gauge boson decaying into the standard model particles through a
kinetic mixing with a $U(1)_{B-L}$ gauge boson; the smallness of
quark's quantum number under the $U(1)_{B-L}$ naturally suppresses the
anti-proton production~\cite{Chen:2008yi}. Perhaps the dark matter
particle has a lepton number~\cite{Asaka:2005cn,Chen:2008dh}, or the lepton
number as well as a discrete symmetry, which is responsible for the
longevity of dark matter, may be slightly broken
altogether~\cite{Takayama:2000uz,Buchmuller:2007ui,Ibarra:2008qg}.  The other
category introduces a light particle in the dark sector so that the dark matter
particle annihilates or decays into the light particles, which then
decay into the standard model particles. If the mass of the light
particle is lighter than $1$\,GeV, the hadronic branch will be
suppressed~\cite{Cholis:2008vb}.  In addition, the presence of such
light particle may enhance the annihilation rate to account
for the relatively large positron production rate suggested by the
PAMELA and ATIC/PPB-BETS data.

Interestingly, the initial energy spectrum of electrons and positrons
are quite different in the above two classes of the dark matter
models. Such difference in the initial source spectrum may persist in
the cosmic-ray electron spectrum observed at the Earth. This will be
an important clue to distinguish the dark matter models, since we
expect to measure the energy spectrum more precisely in the near
future.  For instance, the Fermi satellite~\cite{FGST} can measure
electrons with an energy resolution of about $5$\% at $20$ GeV to
$20$\% at $1000$ GeV~\cite{Moiseev:2007js}.  There is also a dedicated
experiment proposed to measure the electron spectrum,
CALET~\cite{Torii:2006qb}, which is an instrument to observe very high
energy electrons and gamma rays on the Japanese Experiment module
Exposure Facility (JEM-EF) of International Space Station (ISS).  The
CALET detector has a sensitivity to electrons from $1$\,GeV to
$10$\,TeV with an energy resolution better than a few \% for energies
greater than 100 GeV. Those measurements will have much more events than the current ATIC/PPB-BETS data.
Thus, those promising cosmic-ray electron
measurements may help us to distinguish different dark matter models
if the ATIC/PPB-BETS excess is indeed due to the dark matter.

In this letter we study the energy spectrum of electrons generated
through the annihilation and/or decay of dark matter, particularly
paying attention to differences in the energy spectra measured at the
solar system for different initial energy spectra. To illustrate our
idea we will consider the following initial energy spectra of the
electrons: (i) monochromatic (ii) flat and (iii) double-peak ones. We
normalize the production rate of the electrons and positrons so as to
account for the ATIC/PPB-BETS anomaly.  As we will see, the three
cases result in quite different energy spectra at the solar system
even after long propagation through the galaxy. We will also discuss
whether we can distinguish different dark matter models in the
experiments such as Fermi and CALET.  The energy resolution is
essential to identify the origin of  the electron and position excess coming from dark
matter. While the discontinuity of the spectrum expected for the
monochromatic electron spectrum can be identified by Fermi, the other
spectra (ii) and (iii) are less prominent, which may leave room for
astrophysical explanation since the electron spectrum from supernova
remnant will also drop significantly with a certain energy cutoff~\footnote{
It was discussed in Ref.~\cite{Hall:2008qu} whether we can distinguish between dark matter and pulsar origins 
with atmospheric Cherenkov Telescopes.
}.  We will however see that the end point of the distribution will be
clearly seen with the resolution of a few \%,  and
that it will be possible to distinguish the two models (ii) and (iii) at more than
$5 \,\sigma$ C.L.  for the expected statistics at CALET.

\section{Cosmic-ray Electron Energy Spectrum}
\label{sec:2}
\subsection{Initial energy spectrum}
\begin{figure}[t]
\includegraphics[scale=0.7]{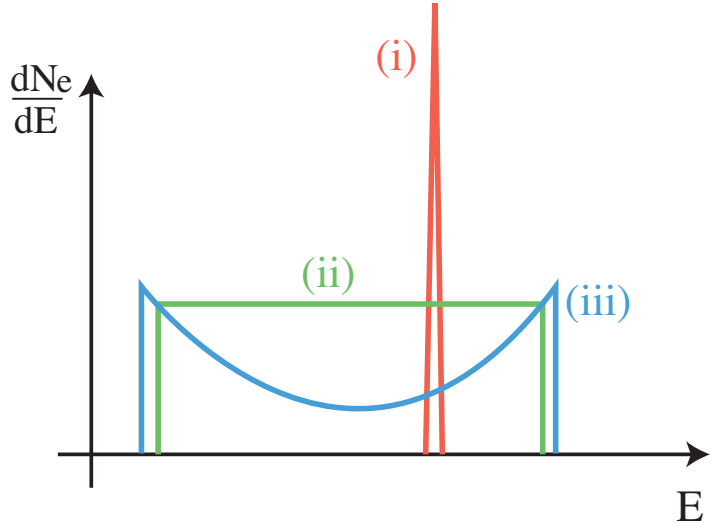}
\caption{Three different initial energy spectra for electrons,
  depending on how the electrons are produced by dark matter.  }
\label{fig:dNdE}
\end{figure}

We consider the following three cases that the initial electron
spectrum is given by (i) monochromatic (ii) flat and (iii) double-peak
ones: (see Fig.~\ref{fig:dNdE})
\bea
\label{eq:ini_spectrum1}
{\rm (i)} \,\,\frac{dN_{e}}{dE} &=& \delta(E - E_{max}),\\
{\rm (ii)} \,\,\frac{dN_{e}}{dE} &=&
\left\{ 
\bear{cc}
1/(E_{max} - E_{min})& {\rm ~~for~~~} E_{min} < E < E_{max}\\
0& {\rm otherwise} 
\eear,
\right.\\
{\rm (iii)} \,\,\frac{dN_{e}}{dE} &=&
\frac{3}{E_{max}^3} \left( \left(E-\frac{E_{max}}{2}\right)^2 + \frac{E_{max}^2}{4} \right)
\theta(E_{max}-E).
\label{eq:ini_spectrum}
\eea
Here we have simply normalized the spectrum as $\int (dN_e/dE) dE =1$,
since we would like to focus on the shape of energy spectrum.

The monochromatic line spectrum is realized if a dark matter particle
of a mass $m_X = E_{max}$ annihilates directly into an
electron-positron pair.  The second and third spectra are obtained if
a heavy dark matter particle $X$ annihilates into a lighter particle
$Y$ (with $m_X \gg m_Y$), which further decays into an
electron-positron pair: $2 X \rightarrow 2 Y \rightarrow 2(e^- +
e^+)$. If $Y$ is a scalar field, an electron and a positron are
emitted isotropically in the rest frame of $Y$, and we obtain a flat
distribution (ii) in the rest frame of $X$'s.  We will approximate
$E_{min} \simeq 0$ and $E_{max} \simeq m_X$ in the following analysis.
On the other hand, if $Y$ is a massive gauge boson, the decay
distribution may not be spherical.  For example, the wino-like dark
matter in a supersymmetric model may annihilate dominantly into a
transverse mode of $W$~\cite{Barger:2008su}~\footnote{
This is the case if $m_X \gg m_W$ and $\mu \gg M_2$, where $m_X$
is the mass of the dark matter, $m_W$ the $W$ boson mass, $\mu$ the
supersymmetric higgsino mass, and $M_2$ the $SU(2)_L$ gaugino
mass. The amplitude of the decay into the longitudinal mode is
equivalent to the pair annihilation  $W^3 W^3 \rightarrow G^- G^+$
which is suppressed proportional to $1/\mu^2$, where $G^\pm$ is the goldstone boson.
}.  The transverse gauge boson decay distribution is proportional to
$(1+\cos^2 \theta)$ in the rest frame of $Y$, where $\theta$ is an
angle between the direction along which $Y$ is boosted and the
electron momentum. In the rest frame of $X$'s, we then have a
double-peak spectrum like (iii), where we have already approximated
the minimum energy to be $0$ as we did for (ii).  In the decaying dark
matter scenario, the relation between $m_X$ and $E_{max}$ should be
replaced with $m_X = 2E_{max}$.  We will see below that the three
initial spectra exhibit themselves in the energy spectra at the solar
system in a different way.

\subsection{Propagation under the galactic magnetic field}
Let us now estimate the energy spectrum of the electrons at the solar
system~\footnote{Although the antiproton measurement provided a
  certain constraint on the dark matter models, we do not take account
  of the antiprotons in this analysis since they are more sensitive to
  the diffusion parameters.}.  After being produced from the
annihilation or decay of the dark matter, an electron will propagate
through the galactic magnetic field~\footnote{ The propagation of an
  electron and a positron can be treated in the same way, and so, we
  call them collectively as an ``electron" afterwards unless otherwise
  stated.  }.  Since the galactic magnetic fields are tangled, the
motion of electrons are described by a diffusion equation. Neglecting
the convection and annihilation in the disk, the steady state solution
should satisfy
\begin{equation}
\nabla \cdot\left[K(E,\vec{r})\nabla f_{e}\right]+\frac{\partial}{\partial E}\left[b(E,\vec{r})f_{e}\right]
+Q(E,\vec{r}) \;=\; 0,\,\label{eq:e_prop}
\end{equation}
where $f_{e}$ is the electron number density per unit kinetic energy,
$K(E,\vec{r})$ a diffusion coefficient, $b(E,\vec{r})$ the rate of
energy loss, and $Q(E,\vec{r})$ a source term of the electrons.
We will neglect the electron mass since the electrons are ultra-relativistic 
in energies of interest. 

The diffusion zone is taken to be a cylinder with half-height $L = 1
\sim 15$\,kpc and a radius $R=20$\,kpc, and the electron number
density is assumed to vanish at the boundary. For simplicity we assume
that $K$ and $b$ are constant inside the diffusion zone and given by
\bea
K(E) &=& K_0\, \lrfp{E}{E_0}{\delta},\\
b(E)&=&\frac{E^2}{E_0 \tau_E},
\eea
where $E_0 = 1$\,GeV and $\tau_E = 10^{16}$\,sec.  The values of
$\delta$, $K_0$ and $L$ must be chosen in such a way that the
measured B/C ratio is reproduced. In Table~\ref{tab:diffusion_models}
we show three sets of such parameters, M2, MED and M1, which yield the
minimum, median, and maximal flux of electrons,
respectively~\cite{Delahaye:2007fr}.

\begin{table}[t]
\begin{center}
\begin{tabular}{|c|ccc|}
\hline
Models&$\delta$&$K_0$\,[kpc$^2$/Myr]&$~L$\,[kpc]\\
\hline
M2&~0.55 &0.00595 &1\\
MED&~0.70 &0.0112 &4\\
M1&~0.46 &0.0765 & 15\\
\hline
\end{tabular}
\end{center}
\caption{The diffusion model parameters consistent with the B/C ratio, yielding 
the minimum, median and maximal electron fluxes, respectively.
\label{tab:diffusion_models} }
\end{table}%

The source term depends on the dark matter distribution, and it is given by
\beq
Q(E,\vec{r}) \;=\; q\cdot (\rho(\vec{r}))^{p}\cdot \frac{dN_{e}(E)}{dE} 
\eeq
with
\beq
q\;=\;\left\{
\bear{ll}
\ds{ \frac{1}{m_X \tau_X}}&~~~{\rm for~~decay}\\
&\\
\ds{\frac{\la \sigma v \ra}{2 m_X^2}}&~~~{\rm for~~annihilation}
\eear
\right.
\eeq
where $p$ equals to $1(2)$ for the decay (annihilation) of dark
matter, $dN_e/dE$ is the initial energy spectrum (\ref{eq:ini_spectrum1}) - 
(\ref{eq:ini_spectrum}), $\rho(\vec{r})$ denotes the dark matter
distribution in our Galaxy, and $\tau_X$, $m_X$, and $\la \sigma v
\ra$ are the lifetime, mass, annihilation cross section of the dark
matter particle $X$, respectively.  In the following analysis we take
the isothermal distribution~\cite{Bergstrom:1997fj}, which is
expressed in terms of the cylinder coordinate, $\vec{r} = (r\cos\phi,
r\sin \phi, z)$ as
\beq
\rho(r,z) \;=\; \rho_\odot\, \frac{r_c^2 + r_\odot^2}{r_c^2 + (r^2+z^2)},
\label{isothermal}
\eeq
with $r_c = 3.5$\,kpc, where $\rho_\odot = 0.30$\,GeV/cm$^3$ denotes
the local dark matter density, and $r_\odot = 8.5$\,kpc is the
distance of the Sun from the galactic center.  We have numerically
checked that our results are not sensitive to the dark matter
profile. This is because an electron of an energy $E \sim 1$\,TeV
typically loses most of the energy before it travels $1$\,kpc away
from the source, and therefore, the dark matter profile around the
galactic center does not change the local electron spectrum
significantly.

The analytic solution of Eq.~(\ref{eq:e_prop}) with the cylindrical
boundary condition was obtained in Ref.~\cite{Hisano:2005ec}. The
electron number density at the solar system ($r_\odot=8.5$\,kpc and
$z_\odot=0$) can be written as
\beq
f_e(E)\;=\; q \cdot\frac{\tau_E E_0}{E^2} \int_E^{E_{max}} dE^\prime\, 
\frac{dN_e(E^\prime)}{dE^\prime}\,
g\left(\lrfp{E_0}{E}{1-\delta} - \lrfp{E_0}{E^\prime}{1-\delta} \right),
\eeq
where we have defined
\bea
g(x)&=& \sum_{n,m =1}^\infty J_0\left(\zeta_n \frac{r_\odot}{R}\right) \sin\lrf{m\pi}{2}
C_{nm}\, e^{-b_{nm} x},\\
C_{nm} &=& \frac{2}{J_1^2(\zeta_n) \pi} \int_0^1 dy_1 \cdot y_1 \int_{-\pi}^{\pi} dy_2\,
J_0(\zeta_n y_1) \sin\left(\frac{m}{2}(\pi-y_2)\right)\,\left(\rho(Ry_1,\frac{L}{\pi} y_2)\right)^p,\\
b_{nm}&=&\frac{K_0 \tau_E}{1-\delta} \left(\frac{\zeta_n^2}{R^2} + \frac{m^2 \pi^2}{4 L^2}\right).
\eea 
Here $J_0$ and $J_1$ are the zeroth and first Bessel functions,
respectively, and $\zeta_n$ ($n=1,2,\cdots$) denotes the successive
zeros of $J_0(x)$.  We can estimate the electron energy spectrum at
the solar system by substituting the dark matter profile
(\ref{isothermal}) and the initial energy spectrum (\ref{eq:ini_spectrum1}) - 
(\ref{eq:ini_spectrum}).

\begin{figure}[t]
\includegraphics[scale=0.7]{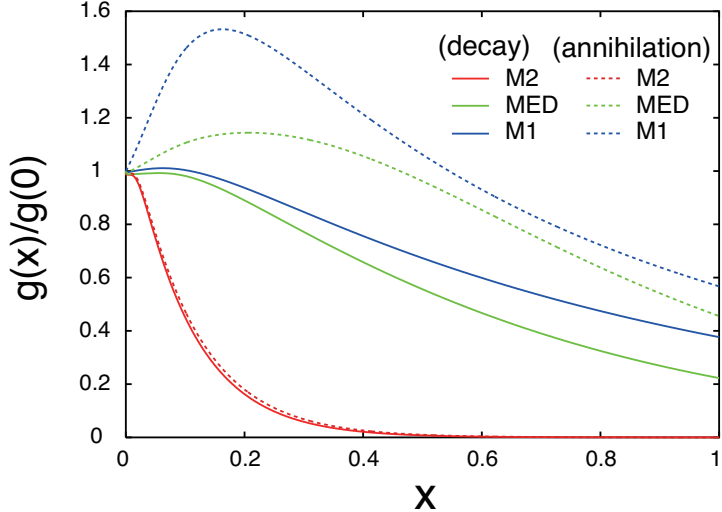}
\caption{The function $g(x)/g(0)$ for decaying (solid) and annihilating (dashed) dark matter for
the M2, MED, and M1 models from bottom to top.}
\label{fig:green}
\end{figure}

In Fig.~\ref{fig:green} we plot $g(x)/g(0)$ as a function of $x$ in
the decaying and annihilating dark matter scenarios for the M2, MED,
and M1 diffusion models.  The function $g(x)$ is a green function,
which expresses a contribution to an electron flux from a distant
source.  We can see that the difference between the decaying and
annihilating dark matter scenarios are almost negligible in the M2
model. This is because the diffusion zone of the M2 model is the
smallest and only the electrons generated in the neighborhood can
reach the Earth. On the other hand, as the diffusion box becomes
larger, the electron tends to travel a longer distance before arriving
at the Earth. In the MED and M1 models, therefore, the form of $g(x)$
is more sensitive to the source distribution, which results in clear
difference between decaying and annihilating dark matter.  Although
one may expect that the resultant energy spectrum would be quite
different between decaying and annihilating dark matter scenarios for
the MED and M1 models, actually this is not the case.  This is
because, as long as we are concerned with the high-end of the electron
flux, it is only
$g(x)$ with $x \ll 1$ that gives most contribution to the flux
$f_e$. Intuitively speaking, since the electrons lose their energies
quickly, the energetic ones must be generated in the neighborhood of
the solar system. This makes it difficult to discriminate the decaying
dark matter scenario from the annihilating one as far as the electron
flux is concerned, since the difference between the two becomes
prominent especially around the galactic center.

\subsection{Electron plus positron energy spectrum}
Let us now estimate the electron spectra at the solar system for
different diffusion models (M2, MED and M1) in the decaying and
annihilating dark matter scenarios with the three different initial
energy spectra (i), (ii) and (iii), using the analytic solution of the
diffusion equation given above.

\begin{figure}[t]
\includegraphics[scale=0.7]{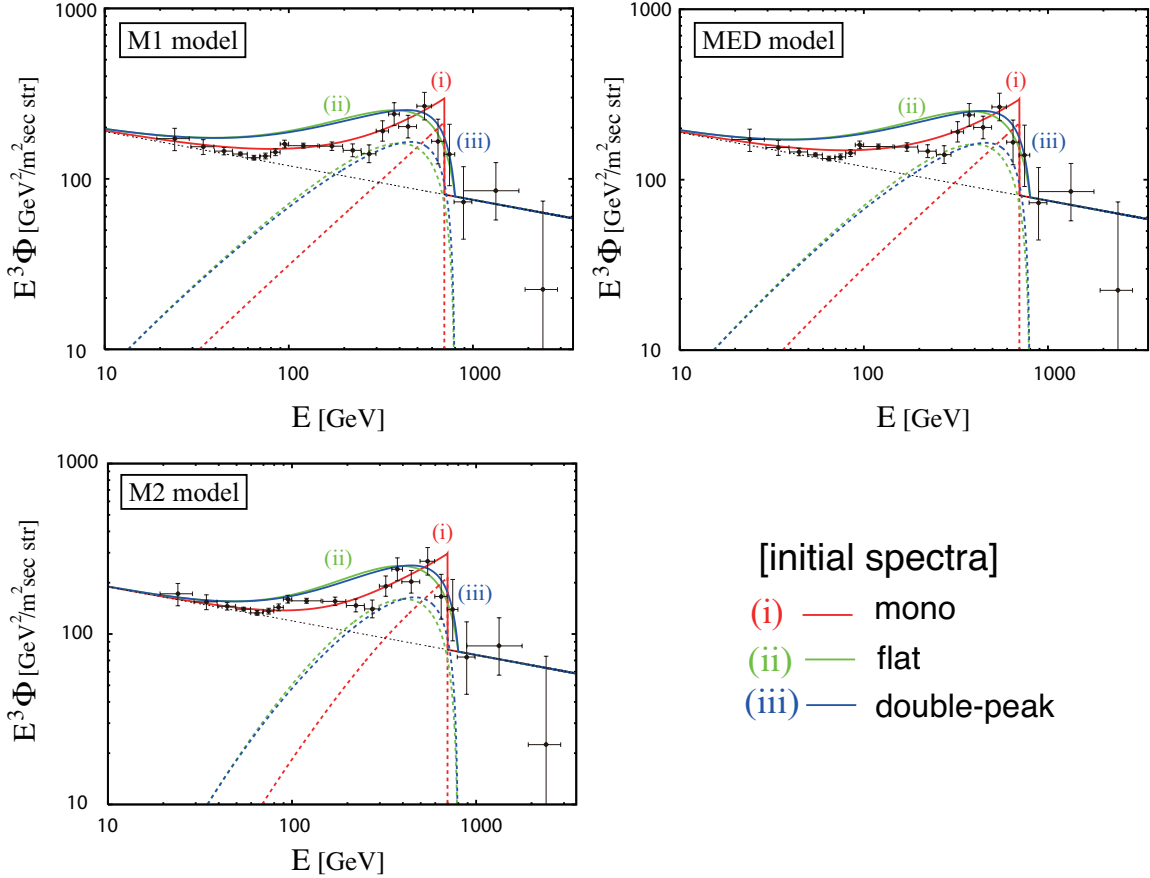}
\caption{The electron plus positron fluxes from the decaying dark
  matter with the three different initial energy spectra, i.e., (i)
  monochromatic, (ii) flat, and (iii) double-peak ones, for the M1,
  MED and M2 diffusion models, together with the ATIC
  data~\cite{ATIC-new}.  The dark matter signal is represented by the
  dotted lines, while the signal plus background is shown as the solid
  lines.}
\label{fig:dec}
\end{figure}

We first show the electron plus positron fluxes (scaled by $E^3$) in
Fig.~\ref{fig:dec} for the decaying dark matter scenario, where we
have assumed the background flux, $\Phi_{bg}(E) = 3 \times 10^2
(E/{\rm GeV})^{-3.2}\, /{\rm(GeV\,m}^2{\rm \,sec\, str)}$.  We adopt the
lifetime and mass of the dark matter $X$ as
\beq
\tau_X\;\simeq\;3.3 \times 10^{26} {\rm\, sec}
{\rm~~ and}
~~m_X = 1400{\rm\, GeV}
\label{eq:taumass}
\eeq
for the monochromatic spectrum, and 
\beq
\tau_X \;\simeq\;1.1 \times 10^{26} {\rm\, sec}
{\rm~~ and}
~~m_X = 1600{\rm\, GeV}
\label{eq:taumass2}
\eeq
for the flat and double-peak ones~\footnote{ Those values are chosen
  for illustration purpose since we are interested in the spectral
  shape.  The best-fitted values of the mass and lifetime should be
  slightly different.  }.  As can be seen from the figure, the three
different initial spectra are clearly reflected in the electron flux
at the solar system, while the dependence on the diffusion models is
rather weak. In particular, the difference between the monochromatic
one (i) and the flat/double-peak ones (ii) and (iii) is significant,
while the latter two (ii) and (iii) look relatively similar. We will
come back to this issue in the next section and study if we can tell
the difference between the flat and double-peak spectra based on the
expected precision of future experiments.

\begin{figure}[t]
\includegraphics[scale=0.7]{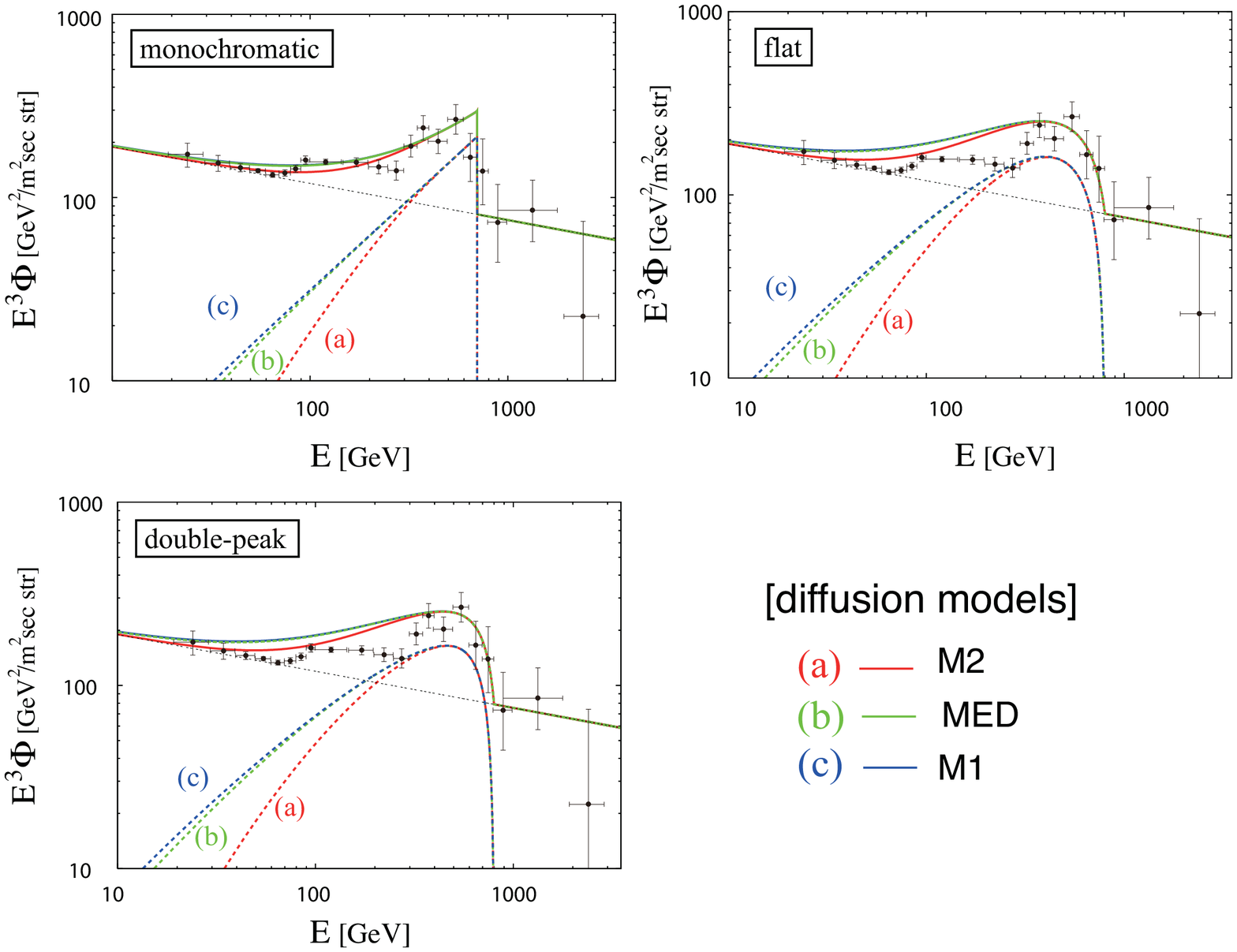}
\caption{The dependence of the electron flux on the diffusion models
  (M2, MED, and M1) in the decaying dark matter scenario with the
  three initial spectra. The dark matter signal is represented by the
  dotted lines, while the signal plus background is shown as the solid
  lines.  }
\label{fig:diff}
\end{figure}

To see how the diffusion models affect the electron flux, we show in
Fig.~\ref{fig:diff} the electron flux for the three diffusion models
in the decaying dark matter scenarios with the three initial
spectra. The parameters are the same as before.  Note that those
features in the electron fluxes are not sensitive to the diffusion
models, especially in the high energy region (say $E \gtrsim (400 -
500)$\,GeV).  In the low energy region, the diffusion modes slightly
affect the electron flux; the M1 and MED models predict a flatter
spectrum than that in the M2 model.  This opens up a possibility to
sort out the dark matter models without suffering an uncertainty as to
the diffusion processes in the galaxy, especially if we focus on the
high-end of the electron flux~\footnote{ In this letter we do not take
  into consideration the electron spectrum below $400$\,GeV, because
  there might be a possible contribution from the nearby
  pulsars~\cite{Hooper:2008kg}, and because the diffusion-model
  dependence may not be negligible.  }.

\begin{figure}[t]
\includegraphics[scale=0.7]{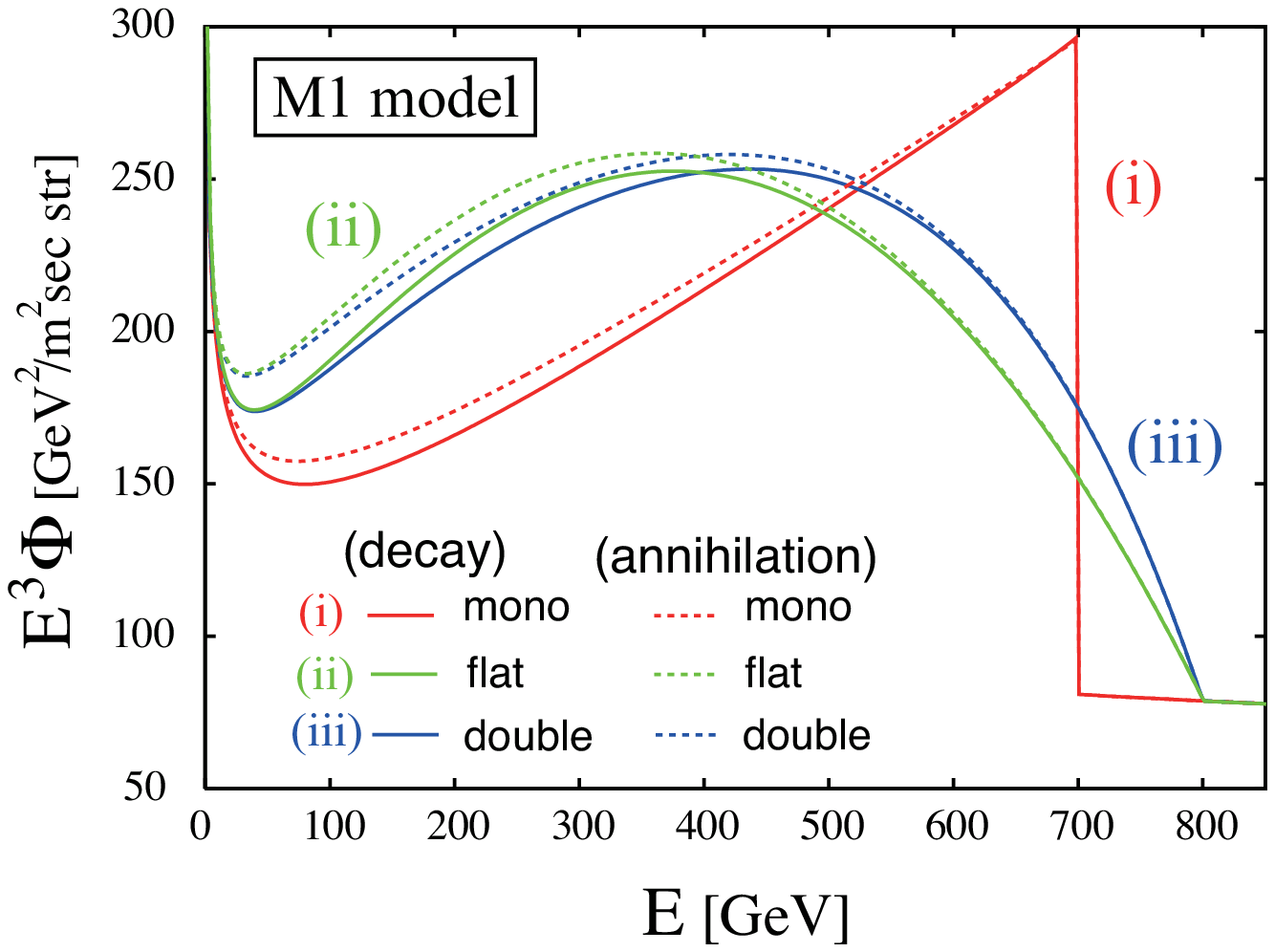}
\includegraphics[scale=0.7]{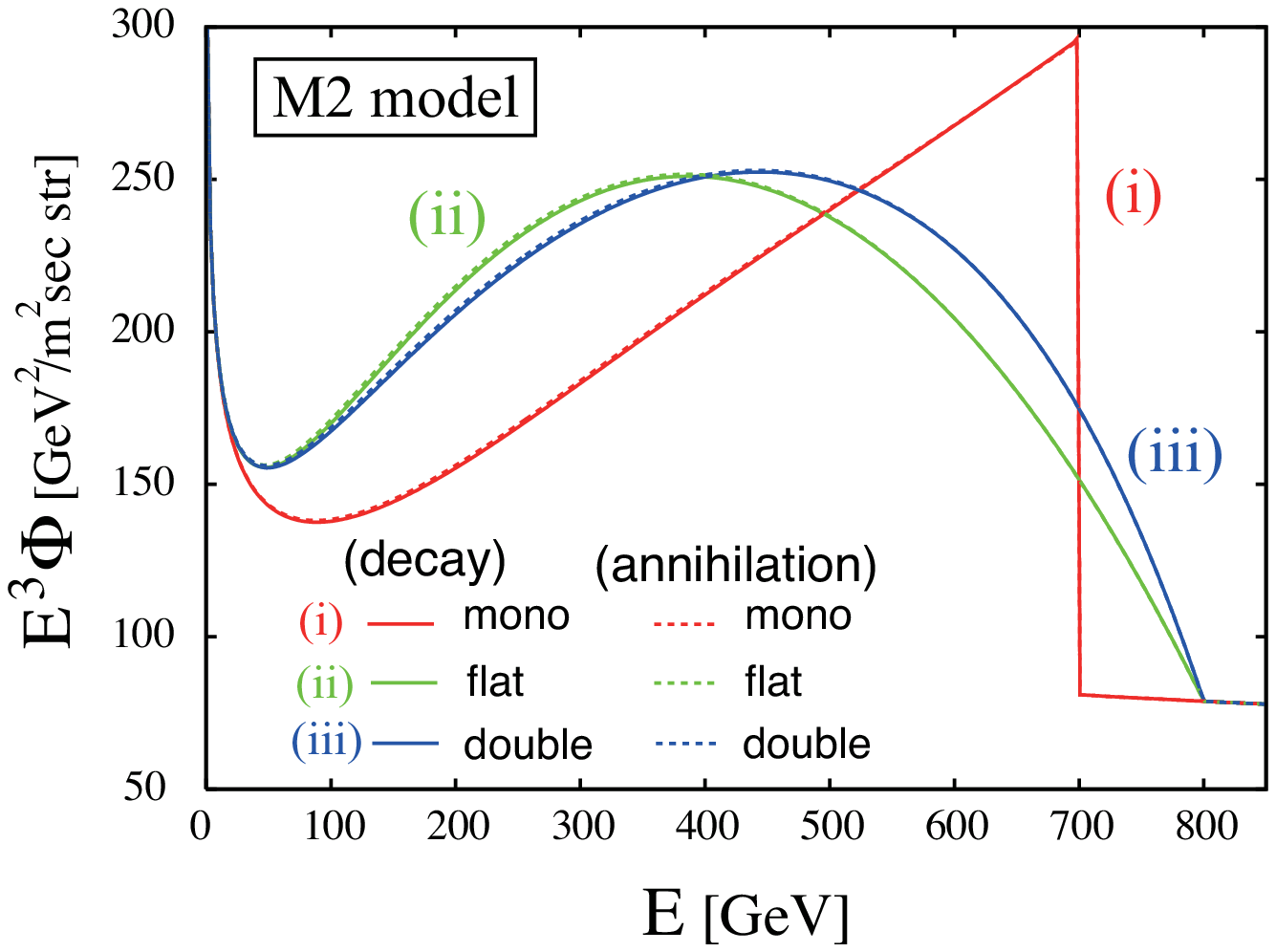}
\caption{The electron plus positron fluxes from the decaying (solid)
  and annihilating (dotted) dark matter with the three different
  initial energy spectra, i.e., (i) monochromatic, (ii) flat, and
  (iii) double-peak ones, for the M2 and M1 diffusion model. The solid
  and dotted lines are quite similar, and almost indistinguishable in
  the M2 model.}
\label{fig:linear}
\end{figure}

Lastly, we show in Fig.~\ref{fig:linear} the electron spectra in both
decaying and annihilating dark matter for the M2 and M1 diffusion
models, where we have set the cross section and the mass as
\beq
\la \sigma v \ra \;=\; 0.7\times 10^{-23}{\rm cm}^3/{\rm sec}
{\rm~~ and}
~~m_X = 700{\rm\, GeV}
\eeq
for the monochromatic spectrum, and 
\beq
\la \sigma v \ra \;=\; 2.4\times 10^{-23}{\rm cm}^3/{\rm sec}
{\rm~~ and}
~~m_X = 800{\rm\, GeV}
\eeq
for the flat and double-peak ones. The mass and lifetime for the decaying dark matter 
are same as before.  As mentioned in the previous subsection, the electron spectra
in the annihilating dark matter scenario look quite similar to those
in the decaying one especially in the M2 model. Although not shown in
the figure, the MED diffusion model is somewhat between the two
models. Also one can more clearly see the difference between the flat
and double-peak spectra in Fig.~\ref{fig:linear}.

\section{Future experiments and dark matter model selection}
In this section we roughly estimate how much statistics
and precision are needed in future experiments in order to tell one
dark matter model from another.  Before proceeding further, however,
it will be useful to briefly review the Fermi and CALET experiments.

In the near future, we expect to measure the energy spectrum of the
cosmic-ray electrons more precisely.  The Fermi satellite~\cite{FGST}
has an $18$ silicon-strip tracker and an $8.5 \,X_0$ thick CsI
calorimeter~\footnote{
$X_0$ denotes the radiation length.
} with a geometric factor about $5 \,{\rm\,m}^2{\rm\,str}$ and the
energy resolution of about $5 - 10$\% over the energy range between
$10$ and $300$\,GeV. The central issue is how to select electrons
while suppressing the hadron (mainly proton) contamination. After
applying a set of selections~\cite{Moiseev:2007js}, the residual
proton contamination is found to be around $3$\% while retaining
almost $30$\% of electrons, and the geometric factor for electrons
turns out to be around $0.8{\rm\,m}^2 {\rm str}$($0.6 {\rm\,m}^2 {\rm str}$) at
$600(800)$\,GeV with an energy resolution of $5$\% at $20$\,GeV to
$20$\,\% at $1000$\,GeV.

\begin{figure}[t]
\includegraphics[scale=0.65]{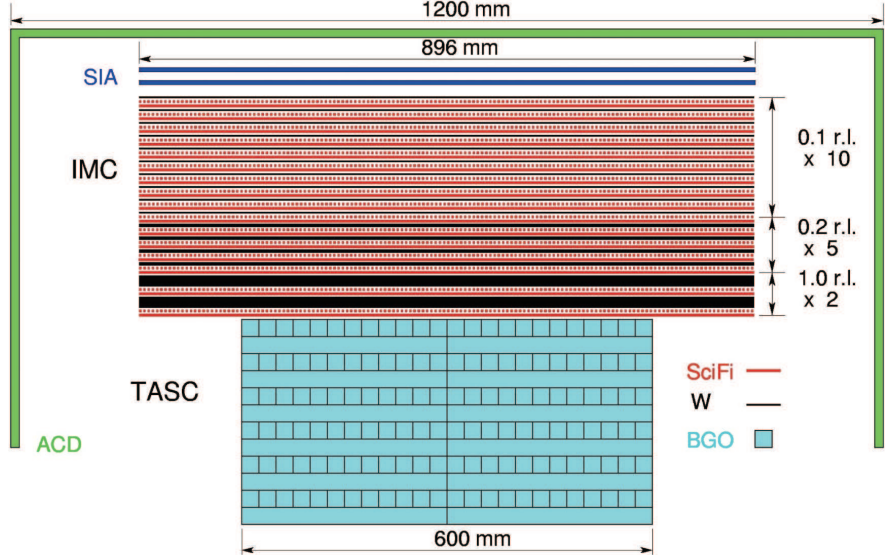}
\caption{Schematic side view of the CALET detector (taken from Ref.~\cite{Torii:2006qb}).
}
\label{fig:calet}
\end{figure}

There is also a dedicated experiment proposed to measure the electron
spectrum; CALET~\cite{Torii:2006qb} is an instrument to observe very
high energy electrons and gamma rays on JEM-EF of ISS.  The CALET
detector, developed based on the previous PPB-BETS balloon
experiments, aims to measure the cosmic-ray electrons from $1$\,GeV to
$10$\,TeV with an energy resolution better than a few \% for energies
greater than 100 GeV.  The CALET detector consists of a combination of
an imaging calorimeter IMC and a total absorption calorimeter
TASC. (See Fig.~\ref{fig:calet}) The geometric factor for electrons is
$0.7{\rm\,m}^2$\,str.  The IMC can achieve the precision necessary to
measure the starting point of an electro-magnetic shower and identify
the incident particle. The TASC measures the development of
electro-magnetic showers to determine the energy of the incident
particle. One of the outstanding features of the CALET detector is
that it achieves $32.8 X_0$ (IMC $+$ TASC), which is large enough to
get rid of the proton contamination efficiently even at an electron
energy of $10$\,TeV.  The energy resolution is estimated to be $7 {\rm
  \%}/ (\sqrt{E/10{\rm GeV)}}$~\cite{Torii:2006qb}.  The experiment
was recently approved for a phase A study aiming at launching the
detector in 2013 for a $5$-year observation.

\begin{figure}[t]
\includegraphics[scale=0.5]{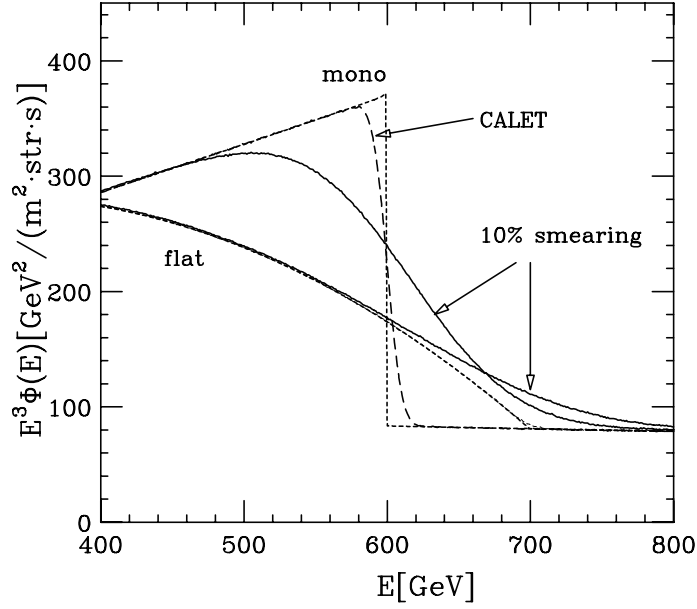}
\caption{The energy spectra for the monochromatic and flat initial spectra before and after
taking account of the energy resolution. The dotted lines are the original ones without smearing;
the solid (long-dashed) lines are obtained after taking account of the smearing effect based on the expected
Fermi (CALET) energy resolution. (The long-dashed and dotted lines are indistinguishable for the flat  initial spectrum.)
}
\label{fig:smear}
\end{figure}

We have seen from Fig.~\ref{fig:dec} that the monochromatic initial
spectrum results in the electron flux with a sharp drop-off. Let us
study whether we can see such a feature in the energy spectrum when
measured with the energy resolution of the Fermi and CALET
experiments. The result is shown in Fig.~\ref{fig:smear}, where we
have set the dark matter mass $m_X = 600 (700)$\,GeV for the
monochromatic(flat) initial spectrum and the cross section $\la \sigma
v \ra = 0.8(2.4) \times 10^{-23}\,{\rm cm}^3/{\rm sec}$. 
We have adopted the annihilating dark matter with the MED diffusion model, 
although the result is not sensitive to the decay/annihilation nor to the diffusion models (see Figs.~\ref{fig:diff} and \ref{fig:linear}).
The dotted lines
are the original ones corresponding to the monochromatic and flat
initial spectra without smearing, while the solid (long-dashed) lines
are obtained after taking account of the smearing effect based on the
expected Fermi (CALET) energy resolution.  
Here we take $10$\, \% energy resolution for Fermi and $(7 {\rm
  }/ (\sqrt{E/10{\rm GeV}})  \oplus 1)$\% for CALET, where $1\%$ stands for
  (unknown) systematic error. 
We can see that, even with
the accuracy of the Fermi satellite, the sharp edge of the energy
spectrum at $E = 600$\,GeV is smeared out, resulting in a smooth
transition from $500$\,GeV to $700$\,GeV (solid), which is less
prominent compared to that expected for CALET (long dashed). 
We expect that the accuracy of the Fermi satellite is good enough to
measure the kinematic structure in the case of the monochromatic spectrum.
 In a case of the flat spectrum, however, the expected spectrum for Fermi is too
broad to extract the dark matter mass, which may leave room for
astrophysical interpretations.

\begin{figure}[t]
\includegraphics[scale=0.7]{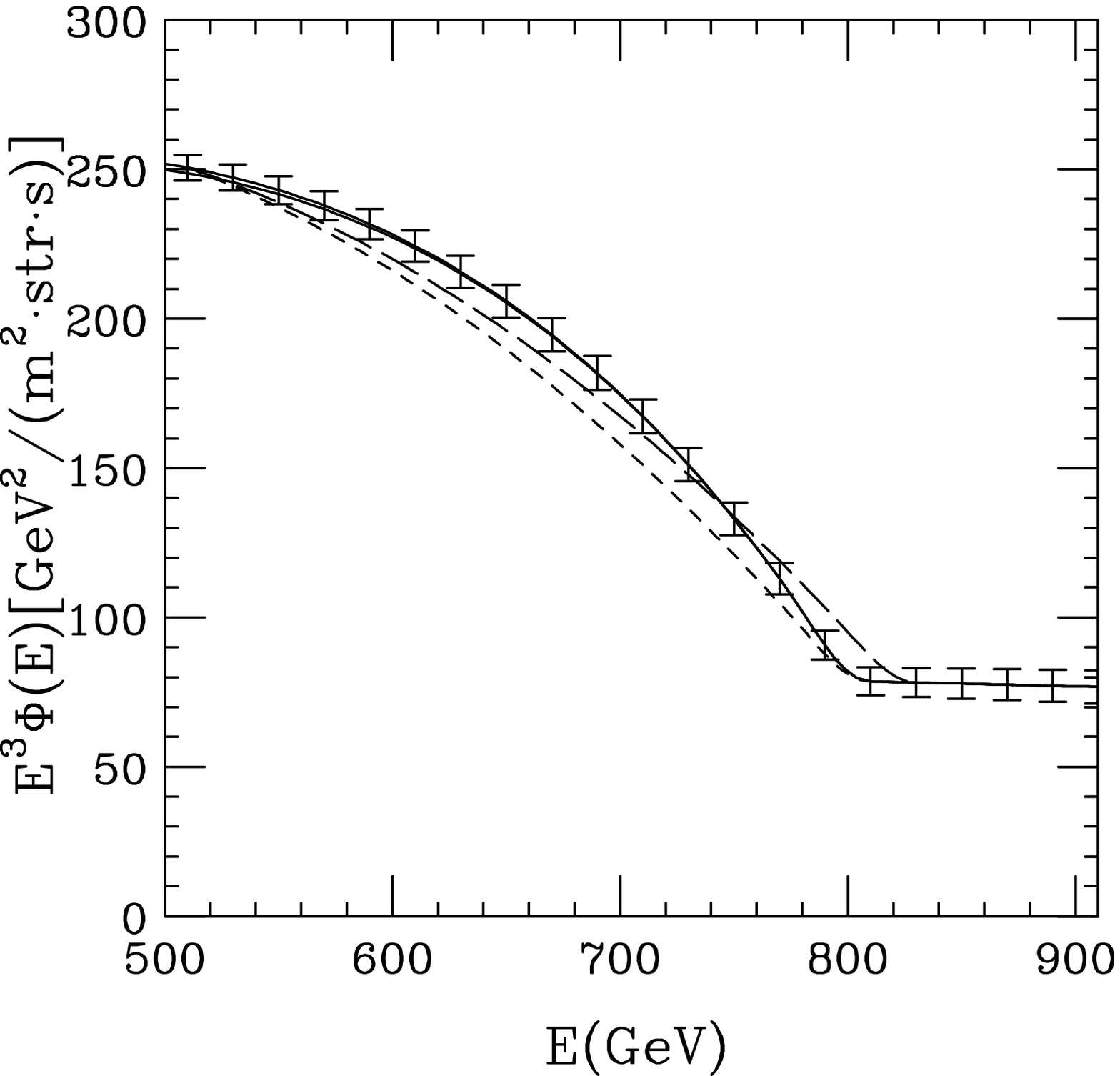}
\caption{The binned $e^- + e^+$ flux together with statistical error shown as bars,
for the double-peak initial spectrum with $m_X = 800$ GeV (solid lines).
The short-dashed and the long-dashed lines correspond to the flat initial spectrum with 
$m_X = 800$ GeV and $m_X = 820$ GeV, respectively. See the text for details.
}
\label{fig:bin}
\end{figure}

Let us now investigate the electron energy distribution for the flat
and double peak initial spectra. From Fig.~\ref{fig:smear} we expect
that CALET can clearly measure the end point of the dark matter
signal.  In Fig.~\ref{fig:bin} we show the expected statistics at
CALET for the annihilating dark matter scenario with the double-peak
initial spectrum (iii), using the MED and M2 diffusion models. We have
adopted the dark matter mass $m_X = 800$\,GeV, and the annihilation
cross section $\la \sigma v \ra = 2.4 \times 10^{-23}{\rm cm}^3/{\rm
  sec}$. Here we assume an exposure of $3\, \,{\rm m}^2 \,{\rm
  str\,years}$, and the errors take into account only the
statistics. To avoid the uncertainty in the diffusion models, we focus
on the flux in the high energy region, i.e., $E > 500$\,GeV. Indeed
the MED and M2 diffusion models do not show any difference in the
plotted region (the upper solid line corresponds to the MED model). 
From the plot, it is obvious that the sharp end of the
distribution will exclude expected astrophysical sources.

We also show the distribution for the flat source spectrum for
$m_X=800$\,GeV (short dashed line). The distribution is normalized so
that it is consistent to the double-peak spectrum at $500$\,GeV by
increasing the pair annihilation cross section. Although the
distribution is quite similar to the solid line, we find more than $1
\sigma$ difference over $11$ bins between $570$\,GeV and 
$770$\,GeV.

Note that the double-peak distribution is an unique signature that
the parent particle is a vector. Although the propagation model we
take in this paper is rather simple, it is tempting to estimate the
$\chi^2$ difference between the two signal profiles
quantitatively. For this purpose, let us define $\delta \chi^2$ as
\beq
 \delta \chi^2  \;=\;    \sum_i \frac{(\phi_{T,i} - \phi_{{\rm flat},i})^2 (\Delta E)^2 \,\Omega^2}{\sigma_i^2},
\eeq
where $\phi_{T,i}$ and $\phi_{{\rm flat},i}$ denote the electron flux for the double-peak
and the flat source spectra, respectively, $\Delta E$ is the width of the energy bin, $\Omega$ is the exposure, 
$\sigma_i$ denotes the standard deviation,
and the summation is taken over the energy bins from $500$\,GeV to $900$\,GeV.
Here we take $\sigma_i^2 = N_i$, where $N_i$ is the number of events
in an $i$-th bin.

We have obtained $\delta \chi^2 \simeq 74.6$ when calculated for
$m_X=800$\,GeV and the flux is normalized so that they are the same
at $E = 500$\,GeV.  If we minimize the $\delta \chi^2$ by varying
$m_X$ with the distribution normalized at 500 GeV, we find the
minimum value of $\delta \chi^2 \simeq 31.8$ for $m_X=820$\,GeV,
corresponding to more than $5 \sigma$ deviations.

\section{Discussion and Conclusions}
\label{sec:4}
In this letter we have estimated the electron energy spectrum at the
solar system for the three different initial energy spectra,
$dN_e/dE$, given by (\ref{eq:ini_spectrum1}) - (\ref{eq:ini_spectrum}), in both decaying and
annihilating dark matter scenarios, varying the diffusion model
parameters.  We have found that the difference in the initial spectra
are reflected in the electron flux measured at the Earth even after
long propagation through the galaxy. We have explicitly shown that
such behavior is robust against changing diffusion parameters as long
as we are concerned with the electron flux in the high energy region
(say $E \gtrsim 400 - 500$\,GeV).  This observation will enable us to
sort out dark matter models by precisely measuring the electron energy
spectrum in a future observation such as CALET or the Fermi satellite
in operation.  We have also shown that the discontinuity predicted by
the monochromatic initial spectrum (i) can be identified with an
energy resolution of Fermi, while the other spectra (ii) and (iii) are
less prominent as the electron spectrum from supernova remnant will
also drop significantly with a certain energy cutoff.  Also we have
studied if we can distinguish the flat and double-peak initial
spectra, which result in relatively similar energy spectra at the
solar system.  We have seen that the end point of the electron
spectrum will be clearly seen with the resolution of about  a few \% assuming
the statistics consistent with the ATIC anomaly, and
that it will be possible to distinguish the two models at more than
$5 \sigma$ C.L.  for the expected statistics at CALET.

\begin{acknowledgments}
We would like to thank Satoshi Shirai and Tsutomu Yanagida for useful discussions.
C.R.C. thanks Institute of Physics, Academia Sinica in Taiwan for its hospitality, 
where part of the work was done.  This work was supported by World Premier International Research Center
Initiative (WPI Initiative), MEXT, Japan. 
\end{acknowledgments}


\begin{thebibliography}{10}
\bibitem{Adriani:2008zr}
  O.~Adriani {\it et al.},
  arXiv:0810.4995 [astro-ph].
  
 \bibitem{ATIC-new}
J.~Chang {\it et al/},
Nature 456 (2008) 362-365.  
  
\bibitem{Torii:2008xu}
  S.~Torii {\it et al.},
  arXiv:0809.0760 [astro-ph].
 

\bibitem{Aharonian(1995)}
  F.~A.~Aharonian, A.~M.~Atoyan and H.~J.~Volk,
  Astron.\ Astrophys.\  {\bf 294}, L41 (1995).
  
\bibitem{Hooper:2008kg}
  D.~Hooper, P.~Blasi and P.~D.~Serpico,
  arXiv:0810.1527 [astro-ph];\\
  H.~Yuksel, M.~D.~Kistler and T.~Stanev,
  arXiv:0810.2784 [astro-ph].

\bibitem{Heinz:2002qj}
  S.~Heinz and R.~A.~Sunyaev,
  Astron.\ Astrophys.\  {\bf 390}, 751 (2002)
  [arXiv:astro-ph/0204183].
  
  

\bibitem{DM-models}
C.~R.~Chen, F.~Takahashi and T.~T.~Yanagida,
 arXiv:0809.0792 {[}hep-ph];
arXiv:0811.0477 [hep-ph];
%
  A.~E.~Nelson and C.~Spitzer,
  arXiv:0810.5167 [hep-ph];
%
  I.~Cholis, D.~P.~Finkbeiner, L.~Goodenough and N.~Weiner,
  arXiv:0810.5344 [astro-ph];
%
  Y.~Nomura and J.~Thaler,
  arXiv:0810.5397 [hep-ph];
%
  R.~Harnik and G.~D.~Kribs,
  arXiv:0810.5557 [hep-ph];
%
  D.~Feldman, Z.~Liu and P.~Nath,
  arXiv:0810.5762 [hep-ph];
 %
  C.~R.~Chen and F.~Takahashi,
  arXiv:0810.4110 [hep-ph];
%
  P.~f.~Yin, Q.~Yuan, J.~Liu, J.~Zhang, X.~j.~Bi and S.~h.~Zhu,
  arXiv:0811.0176 [hep-ph];
%
  Y.~Bai and Z.~Han,
  arXiv:0811.0387 [hep-ph];
%
  P.~J.~Fox and E.~Poppitz,
  arXiv:0811.0399 [hep-ph];
%
  K.~Hamaguchi, E.~Nakamura, S.~Shirai and T.~T.~Yanagida,
  arXiv:0811.0737 [hep-ph];
%
  T.~Hur, H.~S.~Lee and C.~Luhn,
  arXiv:0811.0812 [hep-ph];
%
  E.~Ponton and L.~Randall,
  arXiv:0811.1029 [hep-ph];
%
  M.~Pospelov,
  arXiv:0811.1030 [hep-ph];
  S.~Baek and P.~Ko,
  arXiv:0811.1646 [hep-ph];
%
  E.~J.~Chun and J.~C.~Park,
  arXiv:0812.0308 [hep-ph];
%
  M.~Pospelov and M.~Trott,
  arXiv:0812.0432 [hep-ph];
%
  A.~Arvanitaki, S.~Dimopoulos, S.~Dubovsky, P.~W.~Graham, R.~Harnik and S.~Rajendran,
  arXiv:0812.2075 [hep-ph];
%
  R.~Allahverdi, B.~Dutta, K.~Richardson-McDaniel and Y.~Santoso,
  arXiv:0812.2196 [hep-ph];
%
  K.~Hamaguchi, S.~Shirai and T.~T.~Yanagida,
  arXiv:0812.2374 [hep-ph].


\bibitem{Adriani:2008zq}
  O.~Adriani {\it et al.},
  arXiv:0810.4994 [astro-ph].
   
\bibitem{Yamamoto:2008zz}
  A.~Yamamoto {\it et al.},
  Adv.\ Space Res.\  {\bf 42}, 442 (2008).
   
  
\bibitem{Chen:2008yi} 
C.~R.~Chen, F.~Takahashi and T.~T.~Yanagida,
in Ref.~\cite{DM-models};
 C.~R.~Chen, Mihoko M.~Nojiri,  F.~Takahashi and T.~T.~Yanagida,
 arXiv:0811.3357 [astro-ph].
  
  


\bibitem{Asaka:2005cn} T.~Asaka, K.~Ishiwata and T.~Moroi, 
 Phys.\ Rev.\  D \textbf{73}, 051301 (2006) {[}arXiv:hep-ph/0512118].

\bibitem{Chen:2008dh}
C.~R.~Chen and F.~Takahashi,
arXiv:0810.4110 [hep-ph].
  
  

\bibitem{Takayama:2000uz} F.~Takayama and M.~Yamaguchi, 
 Phys.\ Lett.\  B \textbf{485}, 388 (2000) {[}arXiv:hep-ph/0005214].


\bibitem{Buchmuller:2007ui} 
W.~Buchmuller, L.~Covi, K.~Hamaguchi, A.~Ibarra and T.~Yanagida, 
 JHEP \textbf{0703}, 037 (2007) {[}arXiv:hep-ph/0702184]. 
  
\bibitem{Ibarra:2008qg}
  A.~Ibarra and D.~Tran,
  JCAP {\bf 0807}, 002 (2008)
  [arXiv:0804.4596 [astro-ph]];
  K.~Ishiwata, S.~Matsumoto and T.~Moroi,
  arXiv:0805.1133 [hep-ph];
  

  
\bibitem{Cholis:2008vb}
  I.~Cholis, L.~Goodenough and N.~Weiner,
  arXiv:0802.2922 [astro-ph];
See also
  N.~Arkani-Hamed, D.~P.~Finkbeiner, T.~Slatyer and N.~Weiner,
in Ref.~\cite{DM-models}.










\bibitem{FGST} Fermi Gamma-ray Space Telescope (formerly GLAST) collaboration,
see the webpage: http://fermi.gsfc.nasa.gov/



\bibitem{Moiseev:2007js}
  A.~A.~Moiseev, J.~F.~Ormes and I.~V.~Moskalenko,
  arXiv:0706.0882 [astro-ph].



  
\bibitem{Torii:2006qb}
  S.~Torii  [CALET Collaboration],
  Nucl.\ Phys.\ Proc.\ Suppl.\  {\bf 150}, 345 (2006);
  J.\ Phys.\ Conf.\ Ser.\  {\bf 120}, 062020 (2008).
  
  
 
\bibitem{Hall:2008qu}
  J.~Hall and D.~Hooper,
  arXiv:0811.3362 [astro-ph].

 
 
 
 
\bibitem{Barger:2008su}
 V.~Barger, W.~Y.~Keung, D.~Marfatia and G.~Shaughnessy,
 arXiv:0809.0162 [hep-ph].
 
  
  
\bibitem{Delahaye:2007fr}
  T.~Delahaye, R.~Lineros, F.~Donato, N.~Fornengo and P.~Salati,
  Phys.\ Rev.\  D {\bf 77}, 063527 (2008)
  [arXiv:0712.2312 [astro-ph]].


\bibitem{Bergstrom:1997fj}
  L.~Bergstrom, P.~Ullio and J.~H.~Buckley,
  Astropart.\ Phys.\  {\bf 9}, 137 (1998)
  [arXiv:astro-ph/9712318].

  
\bibitem{Hisano:2005ec}
  J.~Hisano, S.~Matsumoto, O.~Saito and M.~Senami,
  Phys.\ Rev.\  D {\bf 73}, 055004 (2006)
  [arXiv:hep-ph/0511118].

    

\end{thebibliography}
\end{document}